# Does the electromotive force (always) represent work?

C. J. Papachristou[1], A. N. Magoulas[2]

[1]Department of Physical Sciences, Naval Academy of Greece, Piraeus, Greece
E-mail: papachristou@snd.edu.gr

[2]Department of Electrical Engineering, Naval Academy of Greece, Piraeus, Greece
E-mail: aris@snd.edu.gr

**Abstract**

In the literature of Electromagnetism, the electromotive force of a "circuit" is often defined as work done on a unit charge during a complete tour of the latter around the circuit. We explain why this statement cannot be generally regarded as true, although it is indeed true in certain simple cases. Several examples are used to illustrate these points.

## 1. Introduction

In a recent paper [1] the authors suggested a pedagogical approach to the *electromotive force* (emf) of a "circuit", a fundamental concept of Electromagnetism. Rather than defining the emf in an *ad hoc* manner for each particular electrodynamic system, this approach begins with the most general definition of the emf and then specializes to certain cases of physical interest, thus recovering the familiar expressions for the emf.

Among the various examples treated in [1], the case of a simple battery-resistor circuit was of particular interest since, in this case, the emf was shown to be equal to the *work, per unit charge,* done by the source (battery) for a complete tour around the circuit. Now, in the literature of Electrodynamics the emf is often *defined* as work per unit charge. As we explain in this paper, this is not generally true except for special cases, such as the aforementioned one.

In Section 2, we give the general definition of the emf, $\mathcal{E}$, and, separately, that of the work per unit charge, $w$, done by the agencies responsible for the generation and preservation of a current flow in the circuit. We then state the necessary conditions in order for the equality $\mathcal{E}=w$ to hold. We stress that, by their very definitions, $\mathcal{E}$ and $w$ are *different* concepts. Thus, the equation $\mathcal{E}=w$ suggests the possible equality of the *values* of two physical quantities, not the conceptual identification of these quantities!

Section 3 reviews the case of a circuit consisting of a battery connected to a resistive wire, in which case the equality $\mathcal{E}=w$ is indeed valid.

In Sec. 4, we study the problem of a wire moving through a static magnetic field. A particular situation where the equality $\mathcal{E}=w$ is valid is treated in Sec. 5.

Finally, Sec. 6 examines the case of a stationary wire inside a time-varying magnetic field. It is shown that the equality $\mathcal{E}=w$ is satisfied only in the special case where the magnetic field varies linearly with time.

## 2. The general definitions of emf and work per unit charge

Consider a region of space in which an electromagnetic (e/m) field exists. In the most general sense, any *closed* path $C$ (or *loop*) within this region will be called a *"circuit"* (whether or not the whole or parts of $C$ consist of material objects such as wires, resistors, capacitors, batteries, etc.). We *arbitrarily* assign a positive direction of traversing the loop $C$, and we consider an element $\vec{dl}$ of $C$ oriented in the positive direction (Fig. 1).

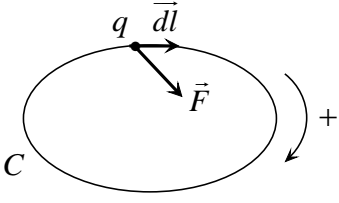

Figure 1

Imagine now a test charge $q$ located at the position of $\vec{dl}$, and let $\vec{F}$ be the force on $q$ at time $t$. This force is exerted by the e/m field itself, as well as, possibly, by additional *energy sources* (e.g., batteries or some external mechanical action) that may contribute to the generation and preservation of a current flow around the loop $C$. The *force per unit charge* at the position of $\vec{dl}$ at time $t$, is

$$\vec{f} = \frac{\vec{F}}{q} \qquad (1)$$

Note that $\vec{f}$ is independent of $q$, since the electromagnetic force on $q$ is proportional to the charge. In particular, reversing the sign of $q$ will have no effect on $\vec{f}$ (although it will change the direction of $\vec{F}$).

In general, neither the shape nor the size of $C$ is required to remain fixed. Moreover, the loop may be in motion relative to an external inertial observer. Thus, for a loop of (possibly) variable shape, size or position in space, we will use the notation $C(t)$ to indicate the state of the curve at time $t$.

We now define the *electromotive force* (emf) of the circuit $C$ at time $t$ as the line integral of $\vec{f}$ along $C$, taken in the *positive* sense of $C$:

$$\mathcal{E}(t) = \oint_{C(t)} \vec{f}(\vec{r},t) \cdot \vec{dl} \qquad (2)$$

(where $\vec{r}$ is the position vector of $\vec{dl}$ relative to the origin of our coordinate system). Note that the sign of the emf is dependent upon our choice of the positive direction of circulation of $C$: by changing this convention, the sign of $\mathcal{E}$ is reversed.

As mentioned above, the force (per unit charge) defined in (1) can be attributed to two factors: the interaction of $q$ with the e/m field itself and the action on $q$ due to any additional energy sources. Eventually, this latter interaction is *electromagnetic* in nature even when it originates from some external mechanical action. We write:

$$\vec{f} = \vec{f}_{em} + \vec{f}_{app} \qquad (3)$$

where $\vec{f}_{em}$ is the force due to the e/m field and $\vec{f}_{app}$ is the *applied force* due to an additional energy source. We note that the force (3) does not include any *resistive* (dissipative) forces that oppose a charge flow along $C$; it only contains forces that may contribute to the generation and preservation of such a flow in the circuit.

Now, suppose we allow *a single charge q* to make a full trip around the circuit $C$ under the action of the force (3). In doing so, the charge describes a curve $C'$ *in space* (not necessarily a closed one!) relative to an external inertial observer. Let $\vec{dl'}$ be an element of $C'$ representing an infinitesimal displacement of $q$ in space, in time $dt$. We define the *work per unit charge* for this complete tour around the circuit by the integral:

$$w = \int_{C'} \vec{f} \cdot \vec{dl'} \qquad (4)$$

For a *stationary* circuit of *fixed* shape, $C'$ coincides with the closed curve $C$ and (4) reduces to

$$w = \oint_{C} \vec{f} \cdot \vec{dl} \qquad (fixed\ C) \qquad (5)$$

It should be noted carefully that the integral (2) is evaluated *at a fixed time t*, while in the integrals (4) and (5) time is allowed to flow! In general, the value of $w$ depends on the time $t_0$ and the point $P_0$ at which $q$ starts its round trip on $C$. Thus, there is a certain ambiguity in the definition of work per unit charge. On the other hand, the ambiguity (so to speak) with respect to the emf is related to the dependence of the latter on time $t$.

The question now is: can the emf be equal *in value* to the work per unit charge, despite the fact that these quantities are defined differently? For the equality $\mathcal{E}=w$ to hold, both $\mathcal{E}$ and $w$ must be defined unambiguously. Thus, $\mathcal{E}$ must be *constant*, independent of time ($d\mathcal{E}/dt=0$) while $w$ must not depend on the initial time $t_0$ or the initial point $P_0$ of the round trip of $q$ on $C$. These requirements are *necessary conditions* in order for the equality $\mathcal{E}=w$ to be meaningful.

In the following sections we illustrate these ideas by means of several examples. As will be seen, the satisfaction of the above-mentioned conditions is the exception rather than the rule!

## 3. A resistive wire connected to a battery

Consider a circuit consisting of an ideal battery (i.e., one with no internal resistance) connected to a metal wire of total resistance $R$ (Fig. 2). As shown in [1] (see also [2]), the emf of the circuit *in the direction of the current* is equal to the voltage $V$ of the battery. Moreover, the emf in this case represents the work, per unit charge, done by the source (battery). Let us review the proof of these statements.

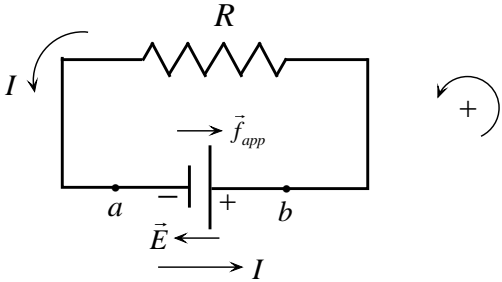

Figure 2

A (*conventionally positive*) moving charge $q$ is subject to two forces around the circuit $C$: an electrostatic force $\vec{F}_e = q\vec{E}$ at every point of $C$ and a force $\vec{F}_{app}$ *inside* the battery, the latter force carrying $q$ from the negative pole $a$ to the positive pole $b$ *through the source*. According to (3), the total force per unit charge is

$$\vec{f} = \vec{f}_e + \vec{f}_{app} = \vec{E} + \vec{f}_{app}$$

The emf in the direction of the current (i.e., counterclockwise), at any time $t$, is

$$\begin{aligned}\mathcal{E} &= \oint_C \vec{f} \cdot \vec{dl} \\ &= \oint_C \vec{E} \cdot \vec{dl} + \oint_C \vec{f}_{app} \cdot \vec{dl} \\ &= \int_a^b \vec{f}_{app} \cdot \vec{dl}\end{aligned} \quad (6)$$

where we have used the facts that $\oint_C \vec{E} \cdot \vec{dl} = 0$ for an electrostatic field and that the action of the source on $q$ is limited to the region between the poles of the battery.

Now, in a steady-state situation ($I$ = constant) the charge $q$ moves at constant speed along the circuit. This means that the total force on $q$ in the direction of the path $C$ is zero. In the interior of the wire, the electrostatic force $\vec{F}_e = q\vec{E}$ is counterbalanced by the resistive force on $q$ due to the collisions of the charge with the positive ions of the metal (as mentioned previously, this latter force does *not* contribute to the emf). In the interior of the (ideal) battery, however, where there is no resistance, the electrostatic force must be counterbalanced by the opposing force exerted by the source. Thus, in the section of the circuit between $a$ and $b$, $\vec{f}_{app} = -\vec{f}_e = -\vec{E}$. By (6), then, we have:

$$\mathcal{E} = -\int_a^b \vec{E} \cdot \vec{dl} = V_b - V_a = V \quad (7)$$

where $V_a$ and $V_b$ are the electrostatic potentials at $a$ and $b$, respectively. We note that the emf is constant in time, as expected in a steady-state situation.

Next, we want to find the work per unit charge for a complete tour around the circuit. To this end, we allow *a single charge q* to make a full trip around $C$ and we use expression (5) (since the wire is stationary and of fixed shape). In applying this relation, time is assumed to flow as $q$ moves along $C$. Given that the situation is static (time-independent), however, time is not really an issue since it doesn't matter at what moment the charge will pass by any given point of $C$. Thus, the integration in (5) will yield the same result (7) as the integration in (6), de-spite the fact that, in the latter case, time was assumed *fixed*. We conclude that the equality $w=\mathcal{E}$ is valid in this case: the emf *does* represent work per unit charge.

## 4. Moving wire inside a static magnetic field

Consider a wire $C$ moving in the $xy$-plane. The shape and/or size of the wire need not remain fixed during its motion. A static magnetic field $\vec{B}(\vec{r})$ is present in the region of space where the wire is moving. For simplicity, we assume that this field is normal to the plane of the wire and directed *into* the page.

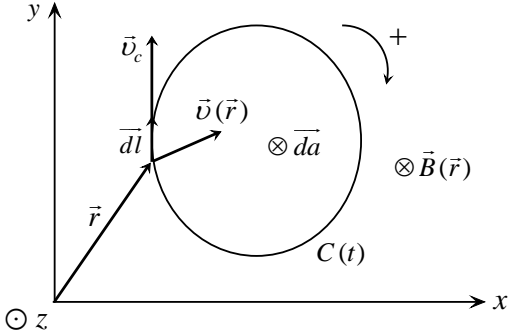

Figure 3

In Fig. 3, the $z$-axis is normal to the plane of the wire and directed towards the reader. We call $\vec{da}$ an infinitesimal normal vector representing an element of the plane surface bounded by the wire (this vector is directed *into* the plane, consistently with the chosen clockwise direction of traversing the loop $C$). If $\hat{u}_z$ is the unit vector on the $z$-axis, then $\vec{da} = -(da)\hat{u}_z$ and $\vec{B} = -B(\vec{r})\hat{u}_z$, where $B(\vec{r}) = |\vec{B}(\vec{r})|$.

Consider an element $\vec{dl}$ of the wire, located at a point with position vector $\vec{r}$ relative to the origin of our inertial frame of reference. Call $\vec{v}(\vec{r})$ the velocity of this element relative to our frame. Let $q$ be a (*conventionally positive*) charge passing by the considered point at time $t$. This charge executes a composite motion, having a velocity $\vec{v}_c$ *along the wire* and acquiring an extra velocity $\vec{v}(\vec{r})$ due to the motion of the wire itself. The total velocity of $q$ relative to us is $\vec{v}_{tot} = \vec{v}_c + \vec{v}$.

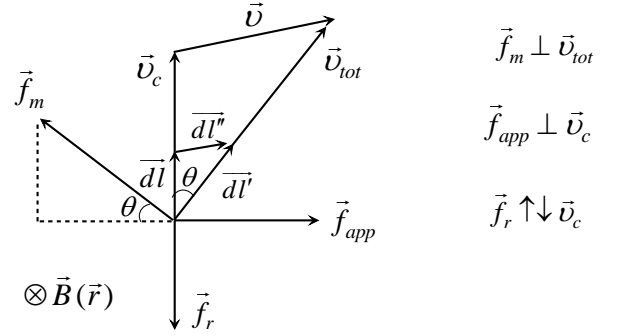

Figure 4

The balance of forces acting on $q$ is shown in the diagram of Fig. 4. The *magnetic force* on $q$ is normal to the charge's total velocity and equal to $\vec{F}_m = q(\vec{v}_{tot} \times \vec{B})$. Hence, the magnetic force per unit charge is $\vec{f}_m = \vec{v}_{tot} \times \vec{B}$. Its component along the wire (i.e., in the direction of $\vec{dl}$) is counterbalanced by the *resistive force* $\vec{f}_r$, which opposes the motion of $q$ along $C$ (this force, as mentioned previously, does *not* contribute to the emf). However, the component of the magnetic force *normal* to the wire will tend to make the wire move "backwards" (in a direction opposing the desired motion of the wire) unless it is counterbalanced by some *external* mechanical action (e.g., our hand, which pulls the wire forward). Now, the charge $q$ takes a share of this action by means of some force transferred to it by the structure of the wire. This force (which will be called an *applied force*) must be *normal* to the wire (in order to counterbalance the normal component of the magnetic force). We denote the applied force per unit charge by $\vec{f}_{app}$. Although this force originates from an external mechanical action, it is delivered to $q$ through an *electromagnetic* interaction with the crystal lattice of the wire (not to be confused with the resistive force, whose role is different!).

According to (3), the total force contributing to the emf of the circuit is $\vec{f} = \vec{f}_m + \vec{f}_{app}$. By (2), the emf at time $t$ is

$$\mathcal{E}(t) = \oint_{C(t)} \vec{f}_m \cdot \overrightarrow{dl} + \oint_{C(t)} \vec{f}_{app} \cdot \overrightarrow{dl}$$

The second integral vanishes since the applied force is normal to the wire element at every point of $C$. The integral of the magnetic force is equal to

$$\oint_C (\vec{\upsilon}_{tot} \times \vec{B}) \cdot \overrightarrow{dl} = \oint_C (\vec{\upsilon}_c \times \vec{B}) \cdot \overrightarrow{dl} + \oint_C (\vec{\upsilon} \times \vec{B}) \cdot \overrightarrow{dl}$$

The first integral on the right vanishes, as can be seen by inspecting Fig. 4. Thus, we finally have:

$$\mathcal{E}(t) = \oint_{C(t)} [\vec{\upsilon}(\vec{r}) \times \vec{B}(\vec{r})] \cdot \overrightarrow{dl} \qquad (8)$$

As shown analytically in [1, 2], the emf of $C$ is equal to

$$\mathcal{E}(t) = -\frac{d}{dt}\Phi_m(t) \qquad (9)$$

where we have introduced the *magnetic flux* through $C$,

$$\Phi_m(t) = \int_{S(t)} \vec{B}(\vec{r}) \cdot \overrightarrow{da} = \int_{S(t)} B(\vec{r}) \, da \qquad (10)$$

[By $S(t)$ we denote *any* open surface bounded by $C$ at time $t$; e.g., the plane surface enclosed by the wire.]

Now, let $C'$ be the path of $q$ in space relative to the external observer, for a full trip of $q$ around the wire (in general, $C'$ will be an *open* curve). According to (4), the work done per unit charge for this trip is

$$w = \int_{C'} \vec{f}_m \cdot \overrightarrow{dl'} + \int_{C'} \vec{f}_{app} \cdot \overrightarrow{dl'}$$

The first integral vanishes (cf. Fig. 4), while for the second one we notice that

$$\vec{f}_{app} \cdot \overrightarrow{dl'} = \vec{f}_{app} \cdot \overrightarrow{dl} + \vec{f}_{app} \cdot \overrightarrow{dl''} = \vec{f}_{app} \cdot \overrightarrow{dl''}$$

(since the applied force is normal to the wire element everywhere; see Fig. 4). Thus we finally have:

$$w = \int_{C'} \vec{f}_{app} \cdot \overrightarrow{dl'} \quad \text{with}$$

$$\vec{f}_{app} \cdot \overrightarrow{dl'} = \vec{f}_{app} \cdot \overrightarrow{dl''} = \vec{f}_{app} \cdot \vec{\upsilon} \, dt \qquad (11)$$

where $\overrightarrow{dl''} = \vec{\upsilon} \, dt$ is the infinitesimal displacement of the wire element in time $dt$.

## 5. An example: Motion inside a uniform magnetic field

Consider a metal bar ($ab$) of length $h$, sliding parallel to itself with constant speed $\upsilon$ on two parallel rails that form part of a U-shaped wire, as shown in Fig. 5. A *uniform* magnetic field $\vec{B}$, pointing into the page, fills the entire region.

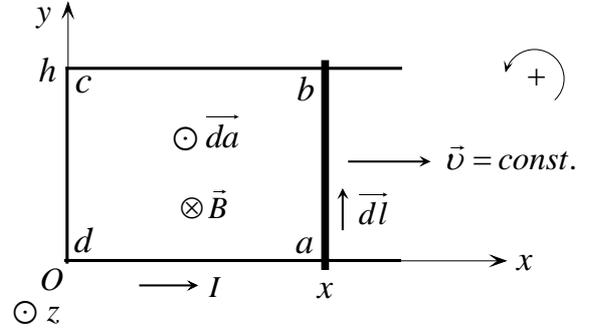

Figure 5

A circuit $C(t)$ of variable size is formed by the rectangular loop ($abcda$). The field and the surface element are written, respectively, as $\vec{B} = -B\hat{u}_z$ (where $B = |\vec{B}| = const.$) and $\overrightarrow{da} = (da)\hat{u}_z$ (note that the direction of traversing the loop $C$ is now counterclockwise).

The general diagram of Fig. 4, representing the balance of forces, reduces to the one shown in Fig. 6. Note that this latter diagram concerns only the *moving* part ($ab$) of the circuit, since it is in this part only that the velocity $\vec{\upsilon}$ and the applied force $\vec{f}_{app}$ are nonzero.

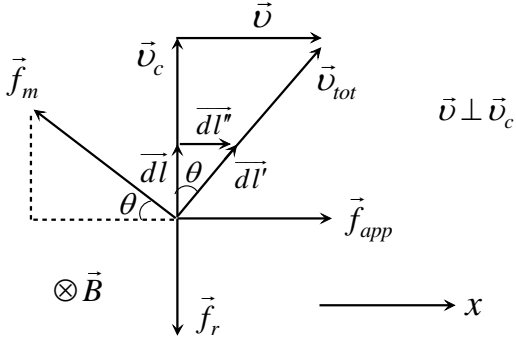

Figure 6

The emf of the circuit at time $t$ is, according to (8),

$$\mathcal{E}(t) = \oint_{C(t)} (\vec{v} \times \vec{B}) \cdot \vec{dl}$$
$$= \int_a^b \upsilon B\, dl = \upsilon B \int_a^b dl = \upsilon B h$$

Alternatively, the magnetic flux through $C$ is

$$\Phi_m(t) = \int_{S(t)} \vec{B}(\vec{r}) \cdot \vec{da} = -\int_{S(t)} B\, da = -B \int_{S(t)} da$$
$$= -Bhx$$

(where $x$ is the momentary position of the bar at time $t$), so that

$$\mathcal{E}(t) = -\frac{d}{dt}\Phi_m(t) = Bh\frac{dx}{dt} = Bh\upsilon$$

We note that the emf is constant (time-independent).

Next, we want to use (11) to evaluate the work per unit charge for a complete tour of a charge around $C$. Since the applied force is non-zero only on the section $(ab)$ of $C$, the path of integration, $C'$ (which is a straight line, given that the charge moves at constant velocity in space) will correspond to the motion of the charge along the metal bar only, i.e., from $a$ to $b$. (Since the bar is being displaced in space while the charge is traveling along it, the line $C'$ will *not* be parallel to the bar.) According to (11),

$$w = \int_{C'} \vec{f}_{app} \cdot \vec{dl'} \quad \text{with}$$
$$\vec{f}_{app} \cdot \vec{dl'} = \vec{f}_{app} \cdot \vec{dl''} = f_{app}\, dl'' = f_{app}\, \upsilon\, dt$$

(cf. Fig. 6). Now, the role of the applied force is to counterbalance the $x$-component of the magnetic force in order that the bar may move at constant speed in the $x$ direction. Thus,

$$f_{app} = f_m \cos\theta = \upsilon_{tot} B \cos\theta = B\upsilon_c$$

and

$$f_{app}\, \upsilon\, dt = B\upsilon\upsilon_c\, dt = B\upsilon\, dl$$

(since $\upsilon_c\, dt$ represents an elementary displacement $dl$ of the charge along the metal bar in time $dt$). We finally have:

$$w = \int_a^b B\upsilon\, dl = B\upsilon \int_a^b dl = B\upsilon h$$

We note that, in this specific example, the value of the work per unit charge is equal to that of the emf, both these quantities being constant and unambiguously defined. This would *not* have been the case, however, if the magnetic field were *nonuniform*!

## 6. Stationary wire inside a time-varying magnetic field

Our final example concerns a *stationary* wire $C$ inside a *time-varying* magnetic field of the form $\vec{B}(\vec{r},t) = -B(\vec{r},t)\hat{u}_z$ (where $B(\vec{r},t) = |\vec{B}(\vec{r},t)|$), as shown in Fig. 7.

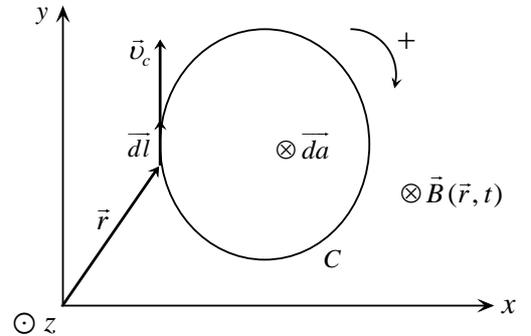

Figure 7

As is well known [1-7], the presence of a time-varying magnetic field implies the presence of an electric field $\vec{E}$ as well, such that Does the electromotive force (always) represent work?

$$\vec{\nabla} \times \vec{E} = -\frac{\partial \vec{B}}{\partial t} \quad (12)$$

As discussed in [1], the emf of the circuit at time $t$ is given by

$$\mathcal{E}(t) = \oint_C \vec{E}(\vec{r},t) \cdot \vec{dl} = -\frac{d}{dt}\Phi_m(t) \quad (13)$$

where

$$\Phi_m(t) = \int_S \vec{B}(\vec{r},t) \cdot \vec{da} = \int_S B(\vec{r},t)\,da \quad (14)$$

is the magnetic flux through $C$ at this time.

On the other hand, the work per unit charge for a full trip around $C$ is given by (5): $w = \oint_C \vec{f} \cdot \vec{dl}$, where $\vec{f} = \vec{f}_{em} = \vec{E} + (\vec{v}_c \times \vec{B})$, so that

$$w = \oint_C \vec{E} \cdot \vec{dl} + \oint_C (\vec{v}_c \times \vec{B}) \cdot \vec{dl}$$

As is easy to see (cf. Fig. 7), the second integral vanishes, thus we are left with

$$w = \oint_C \vec{E} \cdot \vec{dl} \quad (15)$$

The similarity of the integrals in (13) and (15) is deceptive! The integral in (13) is evaluated *at a fixed time t*, while in (15) time is allowed to flow as the charge moves along *C*. Is it, nevertheless, possible that the *values* of these integrals coincide? As mentioned at the end of Sec. 2, a necessary condition for this to be the case is that the two integrations yield time-independent results. In order that $\mathcal{E}$ be time-independent (but nonzero), the magnetic flux (14) – thus the magnetic field itself – must increase *linearly* with time. On the other hand, the integration (15) for $w$ will be time-independent if so is the electric field. By (12), then, the magnetic field must be linearly dependent on time, which brings us back to the previous condition.

As an example, assume that the magnetic field is of the form

$$\vec{B} = -B_0\,t\,\hat{u}_z \quad (B_0 = const.)$$

A possible solution of (12) for $\vec{E}$ is, in cylindrical coordinates,

$$\vec{E} = \frac{B_0 \rho}{2}\hat{u}_\varphi$$

[We assume that these solutions are valid in a limited region of space (e.g., in the interior of a solenoid whose axis coincides with the *z*-axis) so that $\rho$ is finite in the region of interest.] Now, consider a circular wire *C* of radius *R*, centered at the origin of the *xy*-plane. Then, given that $\vec{dl} = -(dl)\hat{u}_\varphi$,

$$\mathcal{E} = \oint_C \vec{E} \cdot \vec{dl} = -\frac{B_0 R}{2} \oint_C dl = -B_0 \pi R^2$$

Alternatively, $\Phi_m = \int_S B\,da = B_0 \pi R^2 t$, so that $\mathcal{E} = -d\Phi_m/dt = -B_0 \pi R^2$. We anticipate that, due to the time constancy of the electric field, the same result will be found for the work $w$ by using (15).

## 7. Concluding remarks

No single, universally accepted definition of the emf seems to exist in the literature of Electromagnetism. The definition given in this article (as well as in [1]) comes close to those of [2] and [3]. In particular, by using an example similar to that of Sec. 5 in this paper, Griffiths [2] makes a clear distinction between the concepts of emf and work per unit charge. In [4] and [5] (as well as in numerous other textbooks) the emf is identified with work per unit charge, in general, while in [6] and [7] it is defined as a closed line integral of the non-conservative part of the electric field that accompanies a time-varying magnetic flux.

The balance of forces and the origin of work in a conducting circuit moving through a magnetic field are nicely discussed in [2, 8, 9]. An interesting approach to the relation between work and emf, utilizing the concept of virtual work, is described in [10].

Of course, the list of references cited above is by no means exhaustive. It only serves to illustrate the diversity of ideas concerning the concept of the emf. The subtleties inherent in this concept make it an interesting subject of study for both the researcher and the advanced student of classical Electrodynamics.